# Outlier detection from ETL Execution trace

Samiran Ghosh
A.K.Choudhury School of
Information Technology
Kolkata-700009, India
Samiranghosh@gmail.com

Saptarsi Goswami
A.K.Choudhury School of
Information Technology
Kolkata-700009, India
saptarsi007@gmail.com

Amlan Chakrabarti
A.K.Choudhury School of
Information Technology
Kolkata-700009, India
acakcs@caluniv.ac.in

*Abstract*— **Extract, Transform, Load (ETL) is an integral part of Data Warehousing (DW) implementation. The commercial tools that are used for this purpose captures lot of execution traces in form of various log files with plethora of information. However there has been hardly any initiative where any proactive analyses have been done on the ETL logs to improve their efficiency. In this paper we utilize outlier detection technique to find the processes varying most from the group in terms of execution trace. As our experiment was carried on actual production processes, any outlier we would consider as a signal rather than a noise. To identify the input parameters for the outlier detection algorithm we employ a survey among developer community with varied mix of experience & expertise. We use simple text parsing to extract these features from the logs, as shortlisted from the survey. Subsequently we applied outlier detection technique (Clustering based) on the logs. By this process we reduced our domain of detailed analysis from 500 logs to 44 logs ( 8 Percentage). Among the 5 outlier cluster, 2 of them are genuine concern, while the other 3 figure out because of the huge number of rows involved.**

## I. INTRODUCTION

ETL (Extract, Transform, and Load) layer is one of the most important layers in the Data warehousing (DW) scheme of things. Companies spend billions of dollars in getting clean, unambiguous data in their data warehouse. Ref [1] observes 70% of the effort and time building a data warehouse goes into this extracting, cleaning, conforming, transforming & loading data. The implementation of ETL jobs can be either hand coded or tool based. The tools that have been predominant are Informatica, Data Stage, Ab-Initio, Oracle Warehouse builder (OWB), SQL Server Integration Service (SSIS), Pentaho etc with the first two being the market leader as per Gartner's 2009 magic quadrant (MQ). We keep our focus on the tool based ETLs as most serious commercial implementations are tool based.

One feature common with all these tools are all of them generates elaborate execution traces (logs) very systematically. Depending on the tracing level (this is generally a configurable property) the log sizes vary. Irrespective of the size, this log files represent plethora of information on various dimension of the load. Industry commonly ignores them and it is inspected only reactively when an ETL job fails or a SLA is missed. There has been little or no distinguishable research effort to analyze this log files for knowledge discovery and necessary action. The efficiency of the information pipeline is of paramount importance to any organization and the organization invests heavily in terms of hardware, software or conforming to standards for the same. However application of (Knowledge Discovery in Database) KDD techniques on this already captured information is something, which has not been explored. In this paper we attempt to address this gap, and demonstrate an approach for the same. Our approach leverages this already captured information, extracts relevant metrics and performs a structured analysis on the ETL processes.

The logs are generally quite large (Level of details can be modified) and represent mines of information. They present two sets of problems: what are the most critical features of an ETL log that we should consider? After identification of the features, how do we extract the same information from the log files in an automated manner? For the first problem, we employ a survey among experienced ETL developers. (More details in Section III). For the second one we use a simple text parsing tool based on Visual Basic for Application (VBA). It is to be noted, our entire discussion and results are based on the logs of a particular ETL tool. However as these are very basic features, we should be able to retrieve them from logs generated by other tools also.

The organization of the paper is as follows: Section II briefs the related works in this area, Section III introduces some basic concepts, and Section IV discusses on the attribute selection details. Section V discusses application of outlier detection technique on the processes and the results. Section VI discusses challenges, future course of work and conclusion.

## II. RELATED WORKS:

Our work in this paper has two main facets. First one is the representation of an ETL Log and the second one is the application of outlier detection methodologies on the same. There has not been any referable scientific research work in representing an execution trace of an ETL Log file. [3], [8] Introduces Meta model based formalism for the ETL processes, [5] analyzes spectrum of ETL Activities in terms of taxonomy. [6] Talks about UML extensions for an ETL task; however it does not touch upon principle characteristics of an ETL Log. For the second point, an analogy with process mining [9] cannot be overlooked however the focus of process mining has more to build the process model from the trace, by finding temporal & causal relationship among the tasks. [10], [16] applies KDD technique on execution traces for intrusion detection & network anomaly detection. [17] Discusses on outlier detection for process logs, however the focus is on finding the structural pattern, the relationship among the activities and then finding outlier [2],[4] Enlists different outlier detection mechanisms and their merits and demerits. [15] Focuses on particular outlier detection methodology like clustering or distance based algorithms. Again in our work, we do not compare various outlier detection technique or deep dive in any one of them, rather use one of them against the ETL Logs and analyze the results to establish the feasibility of our approach.

### III. FOUNDATION:

We start by introducing ETL Tasks, outlier & clustering methods.

#### A. ETL Task:

Informally defined these are the set of processes, which are used to perform the following tasks (not limited to)

1. Extract relevant data from the transactional system and load in Staging.
2. Clean, Conform, De-duplicate the data, apply business transformation and load in Enterprise Data Warehouse (EDW).
3. Load from EDW to Dimensional data marts (DM).
4. Load from DMs to the aggregate layer mainly to reduce the information latency to the business users.

As discussed earlier, there are various commercial tools in the market that meticulously captures execution information.

#### B. Outlier

[7] Defines an outlying observation, or outlier, is one that appears to deviate considerably from other members of the sample in which it occurs. Another definition as observed in [11] is, it is an observation that deviates so much from other observations as to arouse suspicion that it was generated by a different mechanism.

There is a general misconception that outliers are result of faulty measurement and impediment to any meaningful knowledge exploration. The following point is noteworthy in [12], that there are many data mining algorithms for detection of outliers and removing them or minimizing their influence. However this can be signals, not necessarily noise. Our outlier detections will focus on capturing these signals. As observed in [13], there are four major categories of outlier detection namely i) distribution based ii) distance based iii) density based and iv) clustering based. We concentrate on a clustering based approach over here.

#### C. Clustering:

[13] Defines clustering or cluster analysis in the following way:-

Cluster analysis is a set of methodologies for automatic classification of samples into a number of groups using a measure of association, so that the samples in one group are similar and samples belonging to different groups are not similar. The input for a system of cluster analysis is a set of samples and a measure of similarity (or dissimilarity) between two samples.

To compare two observations similarity or distance measure is employed. To express this relationship formally we reference [14]. If there are two clusters $C_1$ and $C_2$, if observations $(p_{11}, p_{12}, p_{13}, .....,p_{1n})$ belong to $C_1$ and $(p_{21},p_{22},p_{23},......, p_{2m})$ belongs to $C_2$ then generically $Dist(p_{1i},p_{1j})<Dist(p_{1i},p_{2k})$, Where $i\neq j$ and $1<=i<=n$, $1<=j<=n$ and $1<=k<=m$. Dist is a function which takes two observations and uses any standard distance measure technique and gives the distance between them as output.

Some of the popular algorithms are K-Means, Nearest Neighbor, PAM etc. For larger data sets, sampling based approaches like CLARA or CLARANS are used.

### IV. ATTRIBUTE SELECTION:

Sample survey using questionnaire is one of the age old techniques for collecting information. We use this method to come up with the most important metrics of an ETL Log.

Our survey methodology has been very simple. Firstly, we shortlisted important features of an ETL process from our past experience. The various features that were under consideration were source and target type (mainframe, flat file, relational), number of. source tables, number of target tables, average source row count, average target row count, number of transformations, number of transformation needing Cache, number of look-up and joiner transformations, average number of lookup rows, total run time, average throughput, low precision flag (a binary variable), deadlock retry logic, average index cache size, average data cache size, commit interval and others (users can enter their text). All these options were presented through a free online survey portal with multi-select check boxes. So to summarize our survey consisted of a single question "What are in your opinion most important features of an ETL Log?"

May be indicatively, but the exhaustiveness of the options could be perceived as there were no remarks in the others section. It was sent to 35+ individuals among whom 22 responded. They had varied experience from 2- 15 years of range with major ETL, BI implementation experience across various industries.

The very clear winners were average source rows, average target rows, number of transformations, total run time, average throughout, number of source tables, as depicted in figure 1.

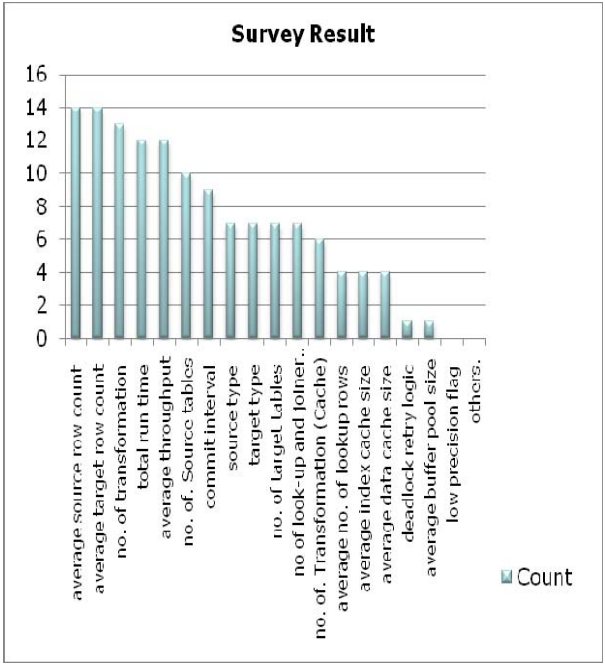

Figure 1: Survey Response.

We used our experience further to shortlist the metrics. As example throughput is a derived metric, from run time and number of rows. So finally we zeroed down on the following attributes: i) number of source rows, ii) number of target Rows iii) total runtime and iv) number of transformations.

Apart from the survey response, another reason for this choice is these parameters are most generic and can be captured form any ETL Process log irrespective of the toolset that is used.

## V. OUR EXPERIMENT AND RESULT

We picked up around 530 + session log from a production environment. Here one point to note is, if we are planning to work with our approach against multiple server logs (i.e. servers with varied hardware and software configuration) or completely different kind of projects (In some project the processes handle complex transformation, some employ pushdown optimization, some handle the processing logic using database scripts) then we need to have a judicious mix of logs, as otherwise because of these reasons, some valid logs can be identified as outlier.

We used simple VBA based string comparison over the logs to extract the 4 metrics. Given a set of log files in a particular folder it generates a CSV file with one row for each log file. This has been shown in figure 3 (the name of the log file has been obfuscated as it is confidential info). We show some little portion to illustrate the logic (complete details omitted for brevity). Among the 4 features that we shortlisted, for our experiment set again the number of transformation had very little or no variation, so though we extracted it from the logs, we actually did not use it as an input parameter to the clustering algorithm.

```
If InStr(strContent, strPat2) > 0 And flagsrc = 0 Then
    strRow = Mid(strContent, InStr(1, strContent, "[") + 1,
InStr(1, strContent, "]") - InStr(1, strContent, "[") - 1)
    strWrite = strWrite + "," + strRow
    'tsw.WriteLine (strWrite)
    flagsrc = 1
End If
```

Figure 2: Code Excerpt of the source rows part

Our code surely has the problem of tight coupling of the underlying schema of the log as this will work on the log of a particular version of a particular tool. We touch upon this in the challenges and conclusion section. After the features are duly extracted we get the results in form of a CSV file, below is a snapshot of the same. The log file names have been masked because of non disclosure.

| | A | B | C | D | E |
|---|---|---|---|---|---|
| 1 | Name | Source Row | Target Row | Time | No .of Transformation |
| 2 | X1.log | 159833 | 159833 | 168 | 2 |
| 3 | X2.log | 4719112 | 4719112 | 9278 | 2 |
| 4 | X3.log | 16178 | 16178 | 1423 | 2 |
| 5 | X4.log | 20715 | 20715 | 338 | 2 |
| 6 | X5.log | 494 | 494 | 82 | 2 |
| 7 | X6.log | 160 | 160 | 35 | 2 |

Figure 3: Snapshot of results CSV.

As our objective is not to find any particular clustering algorithm for outlier detection and then customize it or propose a new way of clustering, rather to show applicability of our idea, we decided to use a commercial data mining tool for the same. Once we have the dataset ready we use SQL Server Analysis service (SSAS) with Microsoft clustering algorithm. Just for interested readers other alternatives could be commercial tools like SAS, SPSS (Now a SAP product), ODM or open source tools like Weka, RapidMiner etc. We no way suggest SSAS is the best tool for the same; the choice is more to do with the intuitive GUIs, user-friendliness and our familiarity with the toolset.

There were quite a few logs with zero source row, zero target row and zero execution time. We removed them from the data set for better clustering results. The number of logs after the exercise stands at 400+, which is surely a good population for clustering analysis. We build the clustering model with all default algorithm parameters. The default algorithm picked up was Scalable EM (Expectation maximization), which assumes each cluster to be a Gaussian distribution.

The distribution of the cluster is captured using the following cluster distribution figure 4. The one that are marked in the figure are certainly the outlier clusters with the least population. We also use the cluster viewer and following is the output as illustrated in figure 4. The weight of the lines between the clusters indicates similarity between the clusters in figure 5. The number of clusters generated here is 10 as per the default parameter, so it suffers from the limitation of K-Means or PAM (K Medoids) of merging meaningful patterns. Separate clustering algorithms can be used for the same and we will take it up as a future course of action.

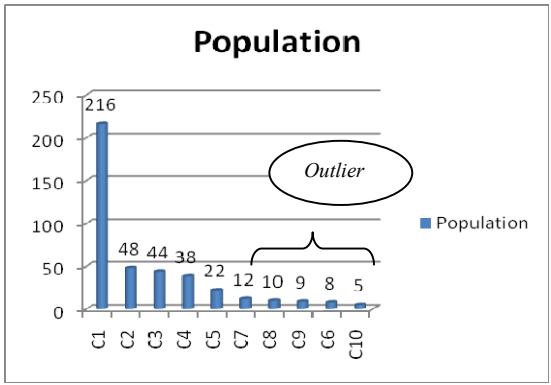

Figure 4: Cluster Population Distribution

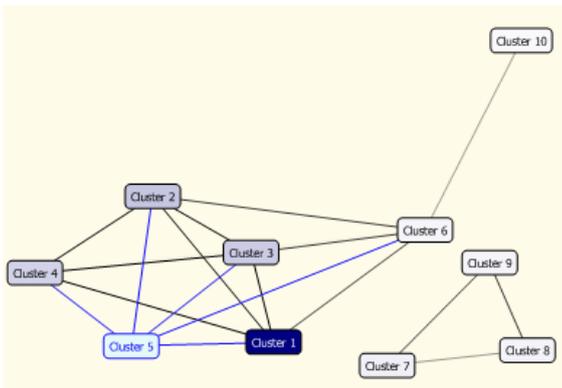

Figure 5: Cluster Diagram

We identify clusters as 'outlier' by an informal guideline of having observations less than 5% of overall population (After the zero metrics logs are removed), which are 20 in this case.

### A.    Analysis of Result

Cluster 10 had 5 members. Upon inspection we found that 3 of them have major issues with query, one of them has a medium issue and the one left had a minor issue. Overall all of them had issue with query like use of functions in where clause, lot of derived table, sub-query and excessive use of union statement. Cluster 6 had 8 observations while 4 of them had major issue with query, 4 of them had an issue with the type of connection that is being used for the data load. Actually instead of a customized external loader connection a generic relational loader has been used. Cluster 7, 8, 9 also had population less than 20.  These are also outliers, however what we see they are outliers because they have huge number of source rows & target rows. So we need not go for a detailed analysis. Also in figure 5 we can see that these clusters have a similarity between themselves.

To summarize, our algorithm helped us to narrow down our investigation scope from 530 to 44, which is 8% of the overall. Between them 2 clusters with minimum population are the actual outliers because of the connection type or the inefficient source query, where as the other 3 are outliers because of huge number of rows than the rest.

However a very important point over here, an outlier does not necessarily indicate a problem as was the case of last 3 clusters. Again a particular source might have too many columns and that is the reason it is taking lot of time for the load as compared with other sources and hence classified as an outlier. So the idea is of an automated indication, which might give some insight. We must also appreciate that this highlights 44 processes to be outlier and hence a detailed investigation becomes possible as against the original 500+ processes.

### VI.    CHALLENGES, FUTURE COURSE & CONCLUSION

The main challenge as discussed earlier is having a universal approach, which can be used against any ETL tool. We have used here a text parsing approach; however most of the vendors allow the log to be exported in XML format conforming to some schema specification. So as a next step we would work on removing the tight coupling. We would propose our program to run periodically and highlight the deviating the ETL Processes, giving immediate benefit in terms of efficiency and effectiveness. Also we would encompass a broader spectrum of developer community to reconfirm our selected metrics. We also need to test out our approach against logs of various projects simultaneously.

Our approach of extracting metrics from ETL log is very simple yet immediate benefits can be achieved from the same. This does not really need any significant additional investment from the organization. All this information is already captured. Neither the text parsing application nor the data mining application involves noticeable cost.

ETL logs generate mines of information. This paper's unique offering is in its attempt to leverage this already captured information in streamlining the processes proactively.